\documentclass[aps,pra,showpacs,notitlepage]{revtex4-1}

\usepackage{graphicx}

\begin{document}




\title{Derivation of quantum mechanics from a single fundamental modification of the relations between physical properties}

\author{Holger F. Hofmann}
\email{hofmann@hiroshima-u.ac.jp}
\affiliation{
Graduate School of Advanced Sciences of Matter, Hiroshima University,
Kagamiyama 1-3-1, Higashi Hiroshima 739-8530, Japan}
\affiliation{JST, CREST, Sanbancho 5, Chiyoda-ku, Tokyo 102-0075, Japan
}

\begin{abstract}
Recent results obtained in quantum measurements indicate that the fundamental relations between three physical properties of a system can be represented by complex conditional probabilities. Here, it is shown that these relations provide a fully deterministic and universally valid framework on which all of quantum mechanics can be based. Specifically, quantum mechanics can be derived by combining the rules of Bayesian probability theory with only a single additional law that explains the phases of complex probabilities. This law, which I introduce here as the law of quantum ergodicity, is based on the observation that the reality of physical properties cannot be separated from the dynamics by which they emerge in measurement interactions. The complex phases are an expression of this inseparability and represent the dynamical structure of transformations between the different properties. In its quantitative form, the law of quantum ergodicity describes a fundamental relation between the ergodic probabilities obtained by dynamical averaging and the deterministic relations between three properties expressed by the complex conditional probabilities. The complete formalism of quantum mechanics can be derived from this one relation, without any axiomatic mathematical assumptions about state vectors or superpositions. It is therefore possible to explain all quantum phenomena as the consequence of a single fundamental law of physics.
\end{abstract}


\maketitle
\section{Introduction}
This paper is an attempt to address the crisis of physics that has emerged with the development of better methods of measurement and control, in particular in the fields of quantum optics and quantum information. The most significant results obtained in these fields are often expressed in the form of paradoxes and have highlighted the fundamental differences between quantum mechanics and our established notions of physical reality. Unfortunately, these results have not led to a better understanding of the physics, but tend to be interpreted as proof that the established formalism cannot be questioned. In fact, there seems to be little doubt left that the relations described by the formalism are correct. However, the difficulties encountered when trying to explain this formalism indicate that it may not be the best formulation of the fundamental physics. It may well be that the actual laws of physics have been obscured and misrepresented by the choice of formulation that emerged from historical accidents. Specifically, we would do well to remember that all formulations carry their own implicit interpretations, so that serious errors of judgment might result from the initial definition of concepts such as ``state'' or ``superposition.'' 

Amid the many new results announced with great fanfare, there is one that would deserve a bit more of our attention because it provides direct experimental evidence of the physics described by states and superpositions. This result is the observation of the wavefunction using the method of weak measurements \cite{Lun11,Aha88}. Significantly, the experimental evidence presented in that work can be understood without any prior knowledge of quantum mechanics, by merely accepting the assumption that a weak measurement can determine a statistical average without any disturbance of the measured system. Nevertheless, the initial explanation of the results was given in terms of the textbook formulation of a quantum state as a superposition of different measurement outcomes. As tempting as it is to simply see these results as a confirmation of established wisdom, one should not overlook that the established idea of superpositions is not connected to any directly observable physics and entered the theory only as a convenient representation of the mathematics. In fact, the impossibility of experimentally observing probability amplitudes appears to be a cornerstone of the Copenhagen interpretation, where superpositions are associated with uncertainties and are consequently treated in the vaguest possible terms. It should therefore come as a bit of a surprise that this abstract mathematical concept appears in the form of a conditional probability in the ``classical'' interpretation of the measurement in \cite{Lun11}.

It has already been pointed out in a large number of works that the statistics observed in weak measurements correspond to a complex valued probability distribution that was already known in the early days of quantum mechanics \cite{Kir33,Dir45,Joh07,Lun12,Hof12}. In particular, it may be said that weak values (i.e. the outcomes of weak measurements) were first discovered theoretically by Dirac, who derived them as a representation of an operator by a function of eigenvalues from two other operator observables \cite{Dir45}. In the context of weak measurements, quantum theory thus implies that the weakly measured property of the system is uniquely determined by the combination of any other two properties, one defined in preparation, and the other in a final measurement. Dirac's algebra of complex conditional probabilities therefore describes the state independent relation between three different properties, and it is this relation that determines the wavefunction of a quantum state observed in a weak measurement according to \cite{Lun11}. 

Amazingly, the relations defined by weak measurements leave no room for uncertainty: once the initial and the final information are combined, the physics of the system is completely determined by the universal relations between these two properties and all other properties of the system. This observation provides the correct explanation for another set of highly publicized results, which demonstrated the prediction of Ozawa that the uncertainty limit of measurements can be overcome by using prior information \cite{Oza03,Hal04,Lun10,Erh12,Roz12,Hof12b}. Putting these pieces of evidence together, it seems that the the role of randomness in quantum mechanics may have been misunderstood. Quantum physics does define universal and deterministic relations between physical properties, where all physical properties can be represented by the eigenvalues observed in precise measurements and the state independent relations between the properties are given by the complex valued conditional probabilities observed in weak measurements. The necessary modification of the classical description then concerns the precise relation of two physical properties to a third, which is the proper expression of universal physical law in quantum mechanics \cite{Hof12,Hof13}.

In the following, I will derive the complete structure of quantum mechanics without ever referring to quantum states or superpositions. This can be achieved by combining the conventional rules of Bayesian probabilities with a single additional law of physics, which describes a fundamental relation between dynamics and statistics and will therefore be referred to as the law of quantum ergodicity. As a result of this law, the phases of complex probabilities can be identified with the action of a reversible transformation between the different properties of a system, where the ratio between the action of transformation and the phase of the complex probabilities is given by Planck's constant. Essentially, quantum mechanics describes the correct relation between the reality of a physical property and the dynamics of its observation, indicating that it is not possible to separate the two. At the most fundamental level, physical properties can only be determined by their observable effects in an interaction with the world of our experience, and this interaction-based definition of physical reality cannot be represented in terms of a scale invariant reduction of geometric shapes to arbitrarily small phase space volumes. Instead, the only universally valid expressions of fundamental relations between different physical properties must be given in terms of complex conditional probabilities, which necessarily replace the geometric shapes used in classical physics. These familiar geometric shapes then emerge only as approximations, in the limit of low resolution, where it is sufficient to identify the gradients of the action-phases of complex probabilities with geometric distances along an approximate trajectory.

\section{Fundamental assumptions of an empirical approach to quantum mechanics}

The following discussion of quantum mechanics is based on the conviction that proper physics proceeds from experimentally observable fact and uses mathematical formalisms only as a tool to efficiently summarize the findings. For this purpose, it is necessary to clarify the assumptions on which the applications of the mathematical tools are based. The problems that have emerged in our understanding of quantum mechanics are caused by the fact that it is not entirely obvious what these assumptions should be. 

From the empirical viewpoint, it is clear that quantum mechanics describes the statistics of measurement outcomes. In this context, it has already been shown that the axioms that describe the mathematical structure of quantum statistics are fundamentally different from the axioms that describe classical statistics \cite{Har01,Fuc01,Gra01,Koc13}. In particular, Hardy \cite{Har01} and Grangier \cite{Gra01} have pointed out that the problem rests with the continuous transformations between discrete observables. However, the proposed axioms merely describe the changes in the mathematical structure, without any direct reference to the physical properties of the system. In particular, it remains unclear how the modified relations between different physical properties relate to the corresponding classical relations. 

Oppositely, it is possible to formulate quantum mechanics operationally, e.g. by referring to a specific set of measurements \cite{Law02,Pat09,Fuc09,Dre13}. Such approaches are usually motivated by the observation that the complete quantum mechanical descriptions of states and measurements can be reconstructed by a sufficiently large set of measurement data. Operational approaches thus provide a consistent description of the relations between different measurements. Moreover, they naturally reproduce the results of classical physics in the limit of low resolution measurements, where statistical averages are sufficient to describe the state of a quantum system. However, the previous operational approaches do not distinguish whether the effects observed in an experiment originate from the system itself, or from the specific circumstances of the measurement setup. Hence it remains unclear how the measurement outcomes relate to the objective properties of a quantum system. 

In the following, the problems of both the axiomatic and the operational approaches are addressed by directly considering the fundamental relations between the different physical properties of a quantum system. For this purpose, the statistical evidence from the measurement of one property of the system must be related to objective properties that can be obtained by performing different measurements. The fundamental assumptions in this approach can be summarized as follows:

\begin{enumerate}
\item Physical systems are described by their observable properties. 
\item There exist universal relations between the physical properties that relate the measurement outcomes of one measurement to the outcomes of other measurements in such a way that statistical predictions based on these relations must always be valid, no matter what the specific situation or circumstances of the experiment may be.
\item Conventional methods of statistical analysis based on conditional probabilities can be used to derive and express the relations between the physical properties, even if the physical properties cannot be observed jointly in any possible experiment.
\end{enumerate}

Significantly, 3. implies that the conventional rules for joint and conditional probabilities also apply to non-commuting physical properties. However, quantum mechanics does not permit joint measurements of such properties. At first sight, this creates a fundamental problem: how can the experimental results be related to joint and conditional probabilities for properties that are not jointly observed in the same experiment?

There have been a number of attempts to address this problem using the tools of quantum state reconstruction. These proposals are rooted in a long history, going back to discussions of the Wigner function as a possible distorted representation of phase space, and to Feynmans various discussions of negative probabilities in quantum mechanics. In more recent times, these ideas have been put into the context of actual experiments, and the results clearly show how non-positive joint probabilities can be recovered from the measurement statistics predicted by the standard quantum formalism \cite{Smi93,Scu94}. In fact, the recent Bayesian approaches to quantum mechanics such as \cite{Law02,Pat09,Fuc09} are mostly motivated by such insights into the relation between experiment and formalism. However, the newer developments on the theoretical side also show that the reconstruction of correlations between separate measurements requires additional assumptions, and in the previous cases (such as the Wigner function and Feynman's negative probabilities), these additional assumptions represent an element of ambiguity in the interpretation of the experimental data. It is therefore difficult to use these specific results as starting points for an empirical reformulation of quantum mechanics. In particular, the previous approaches have not provided an operational definition of Hilbert space vectors in terms of directly observable evidence - which is why I believe that the result reported in \cite{Lun11} may be essential to an empirical understanding of quantum physics.

Taken in the context of these previous approaches, the direct measurements of the wavefunction reported in \cite{Lun11,Lun12} may seem like just another intuitive interpretation of measurement results, similar to the interpretations of \cite{Smi93,Scu94}. It is therefore absolutely necessary to consider the relation between these appoaches in more detail, and I strongly agree that my proposition that the measurement results of \cite{Lun11,Lun12} should be understood as direct evidence of complex conditional probabilities needs to be thoroughly questioned before it can be accepted as the foundation of an empirical approach to quantum mechanics. In particular, it must be understood that (a) the outcomes of weak measurements are correctly predicted by the standard formalism, and (b) the outcomes of weak measurements can be made to fit other statistical models, as seen in the discussion of Bohmian trajectories \cite{Koc11,Sch13}. With respect to (a), the following discussion shows that the outcomes of weak measurements can provide an operational definition of quantum states and quantum coherence that is not provided by the conventional formulation. This means that, although not technically wrong, the standard formulation misleadingly suggests that the mathematics must be accepted without any experimental evidence. Therefore, the direct explanation of weak measurement by complex conditional probabilities may be more valid than the standard explanation by quantum state interferences, just as a description of planetary motion in terms of Kepler orbits is more valid than a technically correct description in terms of epicycles. With respect to (b), it will be important to discuss the experimental procedures used in weak measurements in more detail, and I would like to invite the reader to think about these procedures in terms of the actual physics involved in the measurement interactions. Since I started out as a sceptic with regard to weak measurements, I am well aware that alternative explanations of the experimental results need to be considered. In this context, it may be significant that the statistics discussed in the following can also be obtained by cloning and by measurements at intermediate resolution \cite{Hof12c,Hof12d,Suz12,Bus13}, and that the experimental confirmation of Ozawa uncertainties was achieved without weak measurements \cite{Erh12}. To me, this evidence is convincing enough to conclude that the complex conditional probabilities observed in weak measurements are empirically valid and do not represent an artefact of the specific circumstances created only in weak measurements. Hopefully, the results of the following analysis will motivate a more thorough discussion of the physical effects seen in the various experiments, resulting in a better discrimination between artefacts of our formulations and the actual physics. In particular, I think that the approach in this paper provides a chance to free our thinking from the prejudice that Hilbert space is a necessary assumption in quantum physics, opening the way to a less biased approach to the physical evidence. Most importantly, what follows should be seen as an invitation to a critical discussion of the possible implication of recent developments in quantum measurements, and not as an attempt to claim authority or to monopolize the field of ideas.

\section{Statistical formulation of universal laws of physics}

Ideally, physics should be based on experimental observations obtained and evaluated with only a minimum of theoretical assumptions. We should therefore start by looking at the evidence as it would appear in the readout of our ``classical'' instruments. We can expect with some confidence that any physical property of a system can be measured with arbitrary precision - at least, we have not discovered any fundamental limitations of measurements that relate to a single well-defined property. Moreover, we can say that the result of a measurement may apply equally to the past and to the future, since measurement results can be reproduced in sequential measurements of the same property. Problems only arise when we try to measure different properties jointly. In principle, we can determine any two properties by first measuring one, and then measuring the other. Both results should be equally valid in the time interval between the measurements, and we would expect that measurements performed between initial and final measurement would show how any other physical property of the system depends on the two properties observed in the initial and the final measurement. 

\begin{figure}[th]
\begin{picture}(440,200)
\put(0,0){\makebox(440,200){\vspace*{-3.5cm}
\scalebox{0.7}[0.7]{
\includegraphics{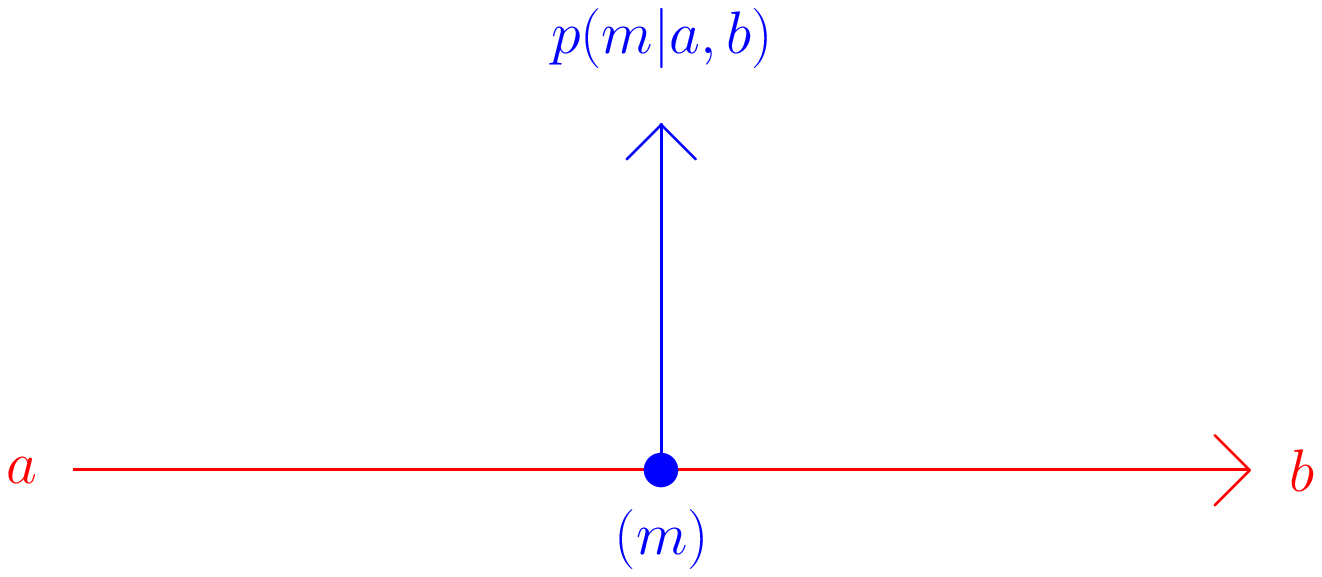}}}}
\end{picture}
\caption{\label{fig1} Illustration of the statistical investigation of probabilities $p(m|a,b)$ conditioned by an initial condition $a$ and a final condition $b$. If the effects of the interaction are negligible, the statistical evidence obtained from the measurement of $m$ allows the experimental evaluation of $p(m|a,b)$.}
\end{figure}

The problem with the ``classical program'' outlined above is that any intermediate measurement involves an interaction, and the dynamics of this interaction may change the value of both the initial and the final property. It is therefore impossible to ignore the effects of dynamics on the measurement results. Although we are tempted to assume that the relations between different properties are independent of the dynamics of transformations, this is not necessarily true: since the reality of a physical property only emerges when the property takes effect in an interaction, it is entirely possible that there is a fundamental relation between transformation dynamics of a system and the effects of the physical properties observed in its measurement. In particular, there is absolutely no reason to require a ``measurement independent reality'' when the only evidence of an objective reality is obtained from interactions with the objects. 

Fortunately, it is possible to analyze the data obtained from intermediate measurements without the assumption of a measurement independent reality. However, this kind of analysis must be based on statistics, since we need to include aspects of the measurement interaction that are beyond our direct control. A particularly clear-cut approach is to make the intermediate measurement interaction weak, so that we can rely on the precise validity of the initial and the final measurement, while obtaining information about the third property by averaging over many trials \cite{Aha88,Lun12,Hof12}. By choosing the right kind of measurements, we can then reconstruct the conditional probabilities for the different measurement outcomes of the third property conditioned by the initial and the final measurement results, as illustrated in Fig. \ref{fig1}. 

Significantly, the results we obtain from the ``classical program'' should represent the fundamental relations between three different properties. The only reason why these relations are formulated in terms of conditional probabilities is that a direct and precise test was found to be impossible. Despite their statistical form, these probabilities should represent well-defined universal relations that apply in any experimental situation, independent of the specific circumstances. It may therefore be useful to identify some formal criteria that can distinguish the statistical expressions of fundamental relations between physical properties from the expressions of randomness usually associated with probability theory. 

Assuming that the relation between three properties is fundamental, the conditional probability $p(m|a,b)$ expresses the dependence of the intermediate property $m$ on the initial property $a$ and the final property $b$. If we wish to consider a fourth property $f$, we only need a relation between $f$ and two of the three properties $a$, $b$, and $m$. Since the relations between three properties are fundamental, the conditional probability $p(f|a,b)$ can be derived from the conditional probability $p(f|m,b)$ using the following chain rule, 
\begin{equation}
\label{eq:chain}
p(f|a,b) = \sum_m p(f|m,b) p(m|a,b).
\end{equation}
In this relation, $f$ is determined by $p(f|m,b)$ without any reference to $a$. It is therefore necessary that knowledge of $a$ does not modify the relation given by $p(f|m,b)$, and that the implications of $a$ for $f$ are already fully accounted for by $m$ and $b$. Thus, the validity of the chain rule in Eq.(\ref{eq:chain}) provides a strong indication that $p(m|a,b)$ is indeed the fundamental relation between $m$, $a$ and $b$.   

A direct test of the fundamental relation $p(m|a,b)$ is obtained if $f=a^\prime$ is either identical to $a$, or represents a different value of the same property, so that $a^\prime \neq a$. In this case, the chain rule requires that the probability of arriving at any value other than the initial value of $a$ is zero,
\begin{equation}
\label{eq:determinism}
\sum_m p(a^\prime|m,b) p(m|a,b) = \delta_{a,a^\prime}.
\end{equation} 
This relation ensures that statements about $(a,b)$ can be converted into equivalent statements about $(m,b)$ without any loss of information. Specifically, the value of $a$ can be uniquely determined by the joint probabilities of $m$ and $b$. In this sense, conditional probabilities that satisfy Eq.(\ref{eq:determinism}) define deterministic relations between $a$ and $m$ under the condition $b$ \cite{Hof12}.  

As mentioned in the introduction, the deterministic relation between initial, final, and intermediate measurement outcomes given by Eq.(\ref{eq:determinism}) can explain the recent results on measurement uncertainties \cite{Oza03,Hal04,Lun10,Erh12,Roz12,Hof12b}. It may therefore be useful to clarify the relation between conditional probabilities and uncertainties. For a probability distribution $p(a)$, the uncertainty of the quantity $A_a$ can be derived by evaluating the differences between two independently obtained samples,
\begin{equation}
\epsilon^2 (A) = \sum_{a,a^\prime} \frac{1}{2}(A_a-A_{a^\prime})^2 p(a) p(a^\prime).
\end{equation}
We can now apply this relation to a situation where the initial condition $b$ is fixed, and the result $m$ is obtained in a final measurement with probability $p(m|b)$. For better comparison with Eq.(\ref{eq:determinism}), it is convenient to replace one of the conditional probabilities of $a$ with a conditional probability of $m$. Assuming that the conditional probabilities originate from the same joint probability of $a$ and $m$ conditioned only by $b$, standard Bayesian rules of probability allow us to convert the probabilities according to 
\begin{equation}
\label{eq:probtrans}
p(m|a,b) p(a|b) = p(a|b,m) p(m|b).
\end{equation}
With this relation, the average conditional uncertainty of $A$ defined by the initial condition $b$ and a final outcome $m$ obtained with a probability of $p(m|b)$ is given by
\begin{equation}
\label{eq:ozawa}
\epsilon^2 (A) = \sum_{a,a^\prime} \frac{1}{2}(A_a-A_{a^\prime})^2 \sum_{m} p(a^\prime|m,b) p(m|a,b) p(a|b).
\end{equation}
If the conditional probabilities satisfy Eq.(\ref{eq:determinism}), the average conditional uncertainty $\epsilon(A)$ is exactly zero, confirming the expectation that there are no random errors in the relation between $(m,b)$ and $a$ described by $p(a|m,b)$.

If there is a set of conditional probabilities that satisfy the relation of Eq.(\ref{eq:determinism}) and are therefore fully deterministic, it is possible to obtain an error free estimate of the value of $A$ based in the initial condition $b$ and the final measurement outcome $m$ by taking the average of $A_a$ for the conditional probability $p(a|m,b)$. Since simultaneous estimates of different properties are possible, there is absolutely no uncertainty limit for joint estimates, or for the relation between estimation errors and the disturbance of another quantity in the measurement interaction. This is the reason why the uncertainty limit for measurements found by Ozawa is much lower than the more familiar limits for quantum states \cite{Oza03}. In fact, it has been pointed out by Hall that the optimal estimate is given by the weak values of the observable \cite{Hal04}, and Lund and Wiseman have shown that Ozawa's definition of measurement errors can be obtained from the complex conditional probabilities observed in weak measurements by assigning complex statistical weights to the differences between the estimate and the eigenvalues \cite{Lun10}. The direct correspondence between the uncertainty of $A$ in Eq.(\ref{eq:ozawa}) and the measurement error of $A$ defined by Ozawa in \cite{Oza03} can be obtained by assuming an error free estimate. In this case, Eq.(\ref{eq:ozawa}) describes the error of the estimate obtained from the averages of $A_a$ for the complex conditional probabilities $p(a|m,b)$ obtained in weak measurements \cite{Hof12b}, which is zero because the complex conditional probabilities are deterministic according to Eq.(\ref{eq:determinism}). The recent experimental confirmations of Ozawa's predictions \cite{Erh12,Roz12} thus provide empirical evidence that the complex conditional probabilities $p(m|a,b)$ observed in weak measurements define the fundamental uncertainty free relation between the three properties $a$, $b$, and $m$. The key to a proper understanding of quantum mechanics is then found in an explanation of the physics described by these complex conditional probabilities. As I shall show in the next section, such an explanation can be given in the form of a single law of physics that defines the relation between dynamics and statistics that is expressed by the complex phases of the probabilities. 

\section{The law of quantum ergodicity}
\label{sec:qergod}

In practical situations, initial conditions only provide partial information about the properties of a system. Nevertheless the precise knowledge of a property $a$ appears to result in a uniquely defined probability distribution $p(b|a)$ for any other property $b$. Since each final measurement of $b$ only reveals a single correct outcome, this probability is an expression of incomplete knowledge of the system and indicates that the value of $b$ is indeed random. It is therefore important to explain why the relative frequencies of the different possibilities $b$ can be given by a uniquely defined probability distribution $p(b|a)$ by identifying the origin of the randomness. 

In analogy to classical statistics, we can find such an explanation in the concept of ergodicity. Specifically, ergodicity relates the dynamics of a system with the expected statistics of a random ensemble by identifying the distribution of various properties with the relative amount of time that the system takes the respective values of these properties during its dynamical evolution. To generalize the ergodic relation to the probabilities $p(b|a)$, we need to consider time evolutions that conserve $a$. The ergodic probability $p(b|a)$ can then be obtained from any complete definition of reality $(a,m)$ by randomizing the dynamics along $a$. Significantly, this kind of randomization corresponds to the effects of the measurement interaction required for a precise measurement of $a$ for an initial condition of $m$. Between preparation $m$ and measurement $a$, the probability of $b$ would be given by $p(b|m,a)$. However, the precise measurement of $a$ will randomize this probability and result in a probability of $p(b|a)$ that is completely independent of $m$. Thus the probabilities $p(b|a)$ are only fundamental in the sense that they are ergodic probabilities derived from dynamic averaging. The randomness that they describe is essentially a randomness of transformations along constant $a$. 

How does quantum mechanics connect the deterministic probabilities $p(m|a,b)$ with the ergodic probabilities $p(b|a)$? In classical physics, the assumption is that the ergodic probability would be obtained by moving along a trajectory of constant $a$, but with varying $b$. The deterministic probability $p(m|a,b)$ would be of little help, since its classical version does not describe the time derivatives of dynamics generated by an energy of $A_a$. However, the results of quantum mechanics suggest that the dynamics of a system play a more fundamental role in the the definition of the deterministic relation between $a$, $b$, and $m$. In the following, it will be shown that the correct quantum mechanical relation between the ergodic probabilities and the conditional probabilities $p(m|a,b)$ is given by a universal law of physics that fundamentally changes the way that the physical properties of a system are related to each other. To emphasize this fundamental modification of the relation between dynamics and statistics that lies at the heart of quantum mechanics, it seems appropriate to refer to this new law of physics as the law of quantum ergodicity. 

A useful starting point for the formulation of the law of quantum ergodicity is the definition of deterministic probabilities in Eq.(\ref{eq:determinism}). For the case of $a=a^\prime$, we find that
\begin{equation}
\label{eq:detsum}
\sum_m p(a|m,b) p(m|a,b) = 1.
\end{equation}  
If we allow only real and positive probabilities, the normalization to $1$ implies that each contribution to the sum is either zero or one, where the single contribution of one indicates the correct value of $m$ determined by this combination of $a$ and $b$. However, quantum paradoxes such as Bell inequality violations clearly show that a simultaneous assignment of $a$, $b$, and $m$ cannot be reconciled with the experimental evidence. It is therefore necessary to modify the relation between $a$, $b$, and $m$ in some fundamental way. As explained above, the available evidence suggests that the correct relations are given by the conditional probabilities observed in weak measurements. Significantly, these conditional probabilities are given by complex numbers, and it has been found that the imaginary parts represent the transformation dynamics of the system \cite{Hof11,Dre12}. It may therefore be possible to obtain the correct fundamental relation between the physical properties of a system by considering complex conditional probabilities, where the complex phases should represent the effects of transformations between the alternative physical properties. For such complex conditional probabilities, the contributions to the sum in Eq.(\ref{eq:detsum}) can have values other than zero or one. In fact, the relation obtained for the complex conditional probabilities observed in weak measurement is particularly simple: the contributions are independent of $b$ and equal to the ergodic probability of $m$ in $a$,
\begin{equation}
\label{eq:ergodicity}
p(a|m,b) p(m|a,b) = p(m|a).
\end{equation}
This relation is the most compact formulation of the law of quantum ergodicity, and all of quantum mechanics can be derived from this single law of physics. 

In the compact form given by Eq.(\ref{eq:ergodicity}), the law of quantum ergodicity states that the absolute value of the contribution of $(m,b)$ to $(a,b)$ in Eq.(\ref{eq:detsum}) is independent of $b$. However, $b$ is still necessary to define a deterministic relation between $a$ and $m$. This problem is solved by the introduction of complex phases. Since $p(m|a)$ is real,
\begin{equation}
\label{eq:arg1}
\mbox{Arg}(p(a|m,b)) = - \mbox{Arg}(p(m|a,b)).
\end{equation}
Here, it is important to observe the importance of the cyclic ordering of $a$, $b$, and $m$. If this order is reversed, the sign of the complex phase must be reversed as well. This is a natural consequence of the connection between the complex phases and the dynamics of a transformation. As explained in \cite{Hof11,Dre12}, the complex phase is related to a force that transforms the initial property $a$ towards the final property $b$. If the role of initial and final property is reversed, the direction of the force is reversed as well, so that
\begin{equation}
\label{eq:arg2}
\mbox{Arg}(p(m|a,b)) = - \mbox{Arg}(p(m|b,a)).
\end{equation}
Comparison between Eq.(\ref{eq:arg1}) and Eq.(\ref{eq:arg2}) shows that the phases only depend on the combination of properties, not on the distinction between target property and conditions. This is consistent with the Bayesian relation between probabilities for different conditions given by Eq.(\ref{eq:probtrans}), where the ratios between the different conditional probabilities are given by the corresponding ergodic probabilities. 

\begin{figure}[th]
\begin{picture}(440,200)
\put(0,0){\makebox(440,200){\vspace*{-3.5cm}
\scalebox{0.7}[0.7]{
\includegraphics{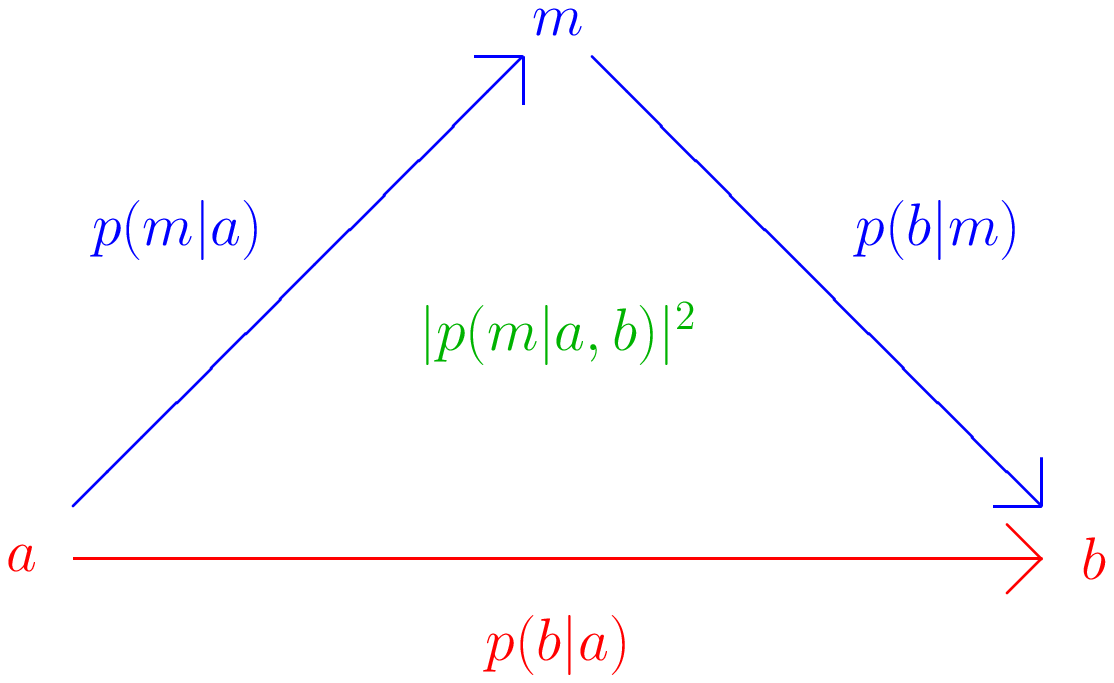}}}}
\end{picture}
\caption{\label{fig2} Illustration of the back-action form of quantum ergodicity. The absolute square of the complex conditional probability defines the ratio of the sequential probabilities $p(b|m)p(m|a)$ and the direct ergodic probability $p(b|a)$.}
\end{figure}

Using the transformations defined by Eq.(\ref{eq:probtrans}), it is possible to formulate an alternative expression of quantum ergodicity, which is more closely related to the problem of measurement back-action. Specifically, the deterministic conditional probability $p(m|a,b)$ assigns a complex probability to $m$ for an initial property $a$ and a final measurement of $b$. However, a measurement of $m$ will disturb the system in such a way that the measurement probability of $b$ for a subsequent measurement changes to the ergodic probability $p(b|m)$. According to Eq.(\ref{eq:ergodicity}) and the Bayesian relation of Eq.(\ref{eq:probtrans}), the probability of first obtaining $m$ and then obtaining $b$ in a subsequent measurement is given by
\begin{equation}
\label{eq:backact}
p(b|m) p(m|a) = p(b|a) |p(m|a,b)|^2.
\end{equation}
This alternative formulation of the law of quantum ergodicity expresses the effect of the dynamical disturbance of the property $b$ in the measurement of $m$. As illustrated in Fig. \ref{fig2}, the absolute value of the complex conditional probability $p(m|a,b)$ is obtained from the ratio between the sequential measurement probability $p(b|m)p(m|a)$ and the direct probability $p(b|a)$, highlighting the relation between complex probability and measurement interaction. Specifically, the back-action eliminates the part of the fundamental relation $p(m|a,b)$ that is expressed by the complex phase, while the absolute value of $|p(m|a,b)|$ is fully described by the ergodic probabilities. Since the measurement back-action corresponds to a randomization of the  dynamics along $m$, this formulation of quantum ergodicity strongly suggests that the dynamics along $m$ can be described in terms of phase shifts for the complex conditional probabilities $p(m|a,b)$.

\section{Transformation distance and action-phases}

In its essence, the law of quantum ergodicity states that the fundamental relations between three physical properties should be expressed by complex conditional probabilities, where the complex phases represent the dynamics of transformations between the properties. The mathematical relation given by Eq.(\ref{eq:ergodicity}) or, equivalently, Eq.(\ref{eq:backact}) provides the specific rule that relates the effects of the properties $a$, $b$, and $m$ to each other. Using this rule, we can now derive the effects of transformations on the fundamental relations between different physical properties. 

According to Eq.(\ref{eq:ergodicity}), the product of the complex conditional probabilities $p(m|a,b)$ and $p(a|m,b)$ does not depend on $b$. It is therefore invariant under reversible transformations of $b$ into $U(b)$. This invariance can also be expressed in terms of the back-action relation in Eq.(\ref{eq:backact}), as
\begin{equation}
|p(m|a,b)|^2 \frac{p(b|a)}{p(b|m)} = |p(m|a,U(b))|^2 \frac{p(U(b)|a)}{p(U(b)|m)}.
\end{equation}
If the transformation $U=U_m$ conserves the property $m$, the ergodic probabilities $p(b|m)$ and $p(U_m(b)|m)$ will be equal and the relation between the complex conditional probabilities simplifies to
\begin{equation}
\label{eq:relate}
|p(m|a,b)|^2 \; p(b|a) = |p(m|a,U_m(b))|^2 \; p(U_m(b)|a).
\end{equation}
This identity shows that the difference between $b$ and its transformation $U_m(b)$ can be described by an $m$-dependent phase shift $\phi_m$. Since all probabilities are normalized to one, the relation between the complex conditional probability with and without the transformation can be written as
\begin{equation}
\label{eq:transform}
p(m|a,U_m(b)) = \frac{p(m|a,b) \exp(i \phi_m)}{\sum_m^\prime p(m^\prime|a,b) \exp(i \phi_{m^\prime})}.
\end{equation}
It is therefore possible to define the effects of a reversible transformation $U_m$ that conserves $m$ entirely in terms of the phase shifts $\phi_m$ that need to be applied to each complex conditional probability of $m$ to transform the final condition $b$ into $U_m(b)$. Together with Eq.(\ref{eq:relate}), these phase shifts then define the change in the ergodic probabilities as
\begin{equation}
\label{eq:dyna1}
p(U_m(b)|a) = p(b|a) | \sum_m p(m|a,b) \exp(i \phi_m)|^2.
\end{equation}
The inverse operation is obtained by simply using the complex conjugate phase factors. Since the application of $U_m^{-1}$ to $b$ is equivalent to the application of $U_m$ to $a$, the effects of the reversible transformation $U_m$on $a$ can be given by
\begin{equation}
\label{eq:dyna2}
p(b|U_m(a)) = p(b|a) |\sum_m p(m|a,b) \exp(-i \phi_m)|^2.
\end{equation}
Complex conditional probabilities thus predict the effects of reversible transformations of $a$ on the output statistics of $b$, based on the phases $\phi_m$ of the transformation $U_m$ \cite{Hof11}.  

\begin{figure}[th]
\begin{picture}(440,200)
\put(0,0){\makebox(440,200){\vspace*{-3.5cm}
\scalebox{0.7}[0.7]{
\includegraphics{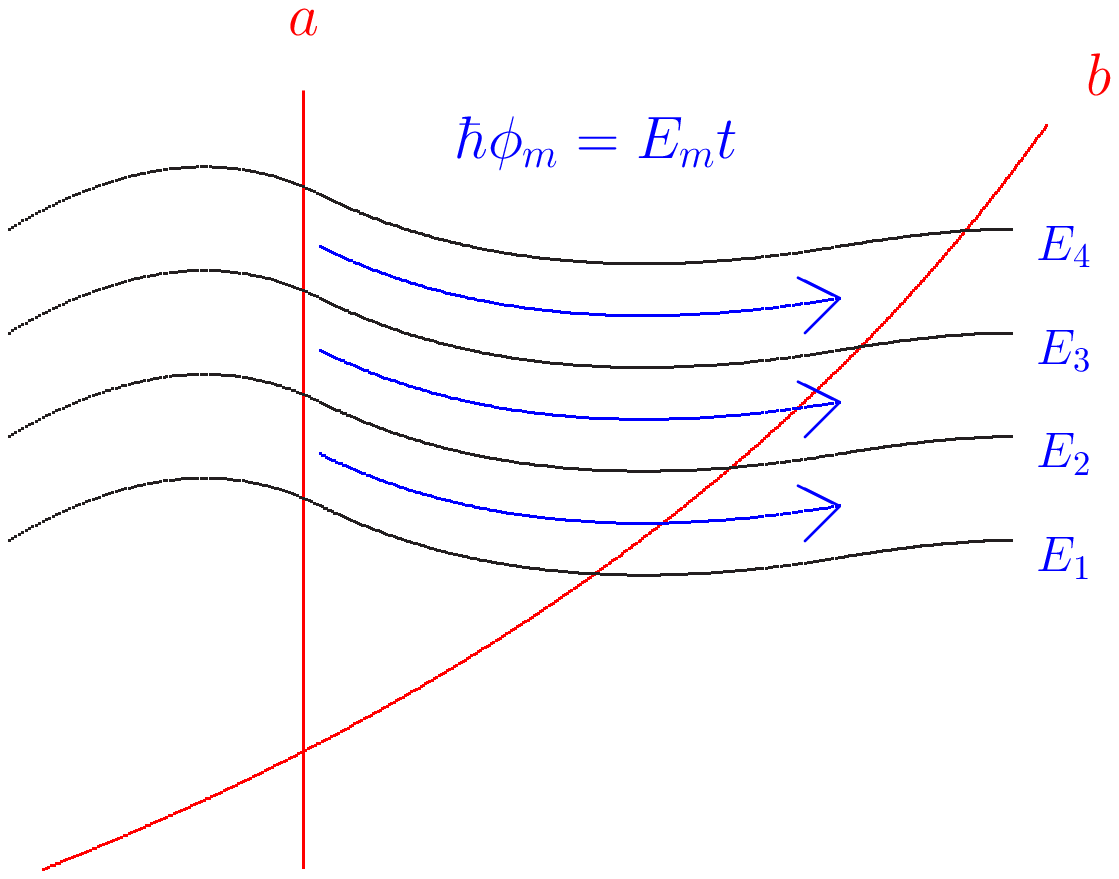}}}}
\end{picture}
\caption{\label{fig3} Schematic illustration of action-phases and transformation distance along constant values of $m$.}
\end{figure}

The derivation given above shows in detail how the law of quantum ergodicity relates the phases of the complex conditional probabilities $p(m|a,b)$ to the action of transformations along constant $m$. Specifically, Eq.(\ref{eq:dyna2}) shows that the probability of obtaining $b$ will be maximal when the phases $\phi_m$ are equal to the phases of the initial complex conditional probabilities $p(m|a,b)$. As discussed in \cite{Hof11}, this means that the complex phases of conditional probabilities describe the transformation distance between $a$ and $b$ along $m$. Since the phases of $p(m|a,b)$ and the phases $\phi_m$ both refer to the action of a transformation along $m$, it may be helpful to refer to them as action-phases to indicate their physical meaning. In fact, the phases $\phi_m$ are closely related to the action in the Hamilton-Jacobi equation of a canonical transformation in classical mechanics, which can be expressed in terms of a product of energy and time. In general, the transformation $U_m$ can be defined in terms of a generator $E_m$ and a conjugate parameter $t$, so that the action-phase $\phi_m$ of the transformation is given by
\begin{equation}
\label{eq:action}
\hbar \phi_m = E_m t.
\end{equation}
Here, the parameter $t$ defines the distance of transformation with respect to the generator $E_m$.
Fig. \ref{fig3} illustrates this role of the generator in the description of transformations between $a$ and $b$ schematically. Since the action is given in terms of a product of energy and time, the fundamental constant $\hbar$ can now be identified as the ratio between the action and the action-phase. It is then possible to explain the relation between the correct quantum mechanical description of physical phenomena and the approximation known as classical physics by identifying the effects of the action-phase at the macroscopic level. 

In the classical limit, a reversible transformation that conserves $E_m$ is described by a trajectory with a constant value of $m$ and a variable value of $b$. The law of quantum ergodicity does not allow such a precise relation between the starting point at $(a,m)$ and the variable $b$. However, it is possible to obtain an approximate relation by coarse graining. In the sum over $m$ that determines the probability of $b$ in Eq.(\ref{eq:dyna2}), the complex values will average to zero if the phases change by more than $2 \pi$ in an interval with an approximately constant absolute value of $p(m|a,b)$. Therefore, the main contributions to $p(b|U_m(a))$ will be found at values of $m$ where the phase gradient in $m$ is close to zero. In the quasi-continuous limit, Eq.(\ref{eq:action}) can be used to express the distance $t$ between $(a,m)$ and $(b,m)$ in terms of the phase gradient of $p(E|a,b)$ for the generator variable $E$,
\begin{equation}
\label{eq:actgrad}
t = \hbar \frac{\partial}{\partial E} \mbox{Arg}(p(E|a,b)).
\end{equation}
The classical limit emerges when the coarse graining in $m$ and in $b$ correspond to an error product of $\delta E \delta t \gg \hbar$. Thus the classical separation of dynamics and the reality of physical properties only emerges when the action enclosed by the error margins is much larger than $\hbar$. At the microscopic level, the physical properties of an object will always be related to each other by complex conditional probabilities, and the transformation distance $t$ must be replaced by the actual complex phases of the conditional probabilities that relate the microscopically precise property $m$ to $a$ and $b$. 

At this point, it is also possible to clarify the origin of quantization itself. If the generator $E_m$ describes a periodic transformation with a period of $T$ in the conjugate parameter $t$, then the transformation defined by $t$ must be equal to the transformation defined by $t+n T$, where $n$ can be any positive or negative integer. Since the action-phases $\phi_m$ that define the transformation depend on $t$ according to Eq.(\ref{eq:action}), this condition can only be satisfied if the differences between adjacent values of $E_m$ are equal to $2 \pi \hbar/T$. Therefore, the law of quantum ergodicity requires that the generators of periodic transformation have quantized values, where the difference between adjacent values is given by the ratio of Planck's constant $2 \pi \hbar$ and the period of the transformation $T$.

\section{Measurement as Interaction}

As discussed in section \ref{sec:qergod}, the formulation of the law of quantum ergodicity given by Eq.(\ref{eq:backact}) can be interpreted directly in terms of the back-action effects of a precise measurement of $m$ between the initial condition $a$ and the final condition $b$. However, back-action can also be interpreted as the effect of an interaction, where the unknown properties of the meter system cause a random transformation of the system. It may therefore be interesting to see if the description of transformation dynamics given by Eq.(\ref{eq:dyna2}) can be used to obtain an expression for the effects of measurement back-action that is consistent with the back-action rule of Eq.(\ref{eq:backact}). 

In order to consider random transformations, it is useful to write out the square of the sum in Eq.(\ref{eq:dyna2}). The relation then reads
\begin{equation}
\label{eq:dyna3}
p(b|U_m(a)) = p(b|a) \sum_{m,m^\prime} p(m|a,b)\; p^*(m^\prime|a,b)\; \exp(-i (\phi_m-\phi_{m^\prime})).
\end{equation}
This formulation clearly shows that the dynamics along $m$ is defined by the differences between the action-phases $\phi_m$ for different values of $m$. It is therefore impossible to describe the dynamics for only a single value of $m$. However, the situation changes when a completely random transformation is considered. In this case, the averages of the phase factors $\exp(-i \phi_m)$ will all be zero, and the sum in Eq.(\ref{eq:dyna3}) reduces to the phase-independent contributions from $m=m^\prime$,
\begin{equation}
\label{eq:random}
p(b|U_{\mathrm{random}}(a)) = p(b|a) \sum_m |p(m|a,b)|^2.
\end{equation}
The right hand side of this equation is a sum of the right hand sides of Eq.(\ref{eq:backact}) over all possible $m$. This result confirms the expectation that a precise measurement of $m$ corresponds to a randomization of the dynamics along $m$, so that the probability of the final measurement outcome $b$ is given by the ergodic probability of $m$, independent of $a$. Specifically, the result of a dynamic randomization along $m$ can be given in terms of the ergodic probabilities $p(m|a)$ and $p(b|m)$ as
\begin{equation}
p(b|U_{\mathrm{random}}(a)) = \sum_m p(b|m)p(m|a).
\end{equation}
It is therefore justified to trace the origin of the back-action to the random dynamics caused by the interaction with the meter system, despite the fact that the form of the back-action given by Eq.(\ref{eq:backact}) refers only to the properties of the system, without any specific reference to the precise form of the interaction. 

The measurement back-action is an essential part of quantum ergodicity, because the complex conditional probabilities that define the relations between three physical properties cannot describe any joint effects of all three properties. The description of transformation dynamics derived from quantum ergodicity shows that such joint effects cannot be observed because the measurement interaction will always result in a randomization of the information expressed by the complex phases of the conditional probabilities. It is therefore reasonable to conclude that the law of quantum ergodicity shows that interaction is a necessary condition of objective reality, and that physical reality cannot be defined in the absence of interactions. 

\section{The origins of Hilbert space}

The theory of complex conditional probabilities developed above represents a complete and consistent formulat‭ion of quantum mechanics. It does not require any of the axioms and postulates usually associated with quantum theory. The central message of this paper is that such concepts are not necessary once the fundamental role of quantum ergodicity is properly understood. However, there is absolutely no contradiction between the standard formulation of quantum mechanics and the one introduced in these pages. In fact, it is now possible to derive the Hilbert space formalism completely from the more fundamental law of quantum ergodicity, providing a physical explanation for concepts that were previously thought to be axiomatic elements of the theory. 

To achieve this derivation of Hilbert space, it is convenient to reformulate the law of quantum ergodicity once more, this time as a relation between the ergodic probability $p(m|a)$ and the absolute square of a re-scaled complex conditional probability,
\begin{equation}
\label{eq:HSform}
p(m|a) = \left| \sqrt{\frac{p(a|b)}{p(m|b)}} p(m|a,b) \right|^2.
\end{equation}
Mathematically, the re-scaled probabilities can be used to define $a$ as a vector of length one in the $d$-dimensional space defined by the $d$ possible values of $m$. Importantly, the property $b$ is necessary to define the phases of the vector components. Quantum ergodicity can thus be used to reduce the role of $b$ to that of a phase standard, so that the relation between $a$ and $m$ under the condition $b$ can be expressed in the form of an inner product of two vectors, $\mid a \rangle$ and $\mid m \rangle$,
 \begin{equation}
\label{eq:amp}
\langle m \mid a \rangle = \sqrt{\frac{p(a|b)}{p(m|b)}} p(m|a,b).
\end{equation}
Thus, the ``state vectors'' of Hilbert space are simply the re-scaled complex conditional probabilities that describe the fundamental relations between the observable properties of a system. Significantly, the ``superposition'' of different values of $m$ arises from the use of a reference $b$, which evaluates $m$ under the condition of a measurement that cannot be performed jointly with the measurement of $m$.  

It is now possible to view the physics of Hilbert space in a new light. Eq.(\ref{eq:amp}) implies that the vector algebra of Hilbert space merely describes the relations between different conditional probabilities. In particular, it is possible to derive the description of an inner product in the $m$-representation from the chain rule of Bayesian probabilities given in Eq.(\ref{eq:chain}). Specifically, the inner product $\langle f \mid a \rangle$ can be expressed as 
\begin{eqnarray}
\label{eq:inner}
\sum_m \langle f \mid m \rangle \langle m \mid a \rangle &=& 
\sqrt{\frac{p(a|b)}{p(f|b)}} \sum_m p(f|m,b) p(m|a,b)
\nonumber \\ &=& \sqrt{\frac{p(a|b)}{p(f|b)}} p(f|a,b).
\end{eqnarray}
Importantly, this sum is responsible for the effects usually interpreted as ``interference'' between the unobserved alternatives $m$. Quantum ergodicity explains that the possibility of expressing the results of one observation in terms of the results of another observation is based on the dynamical relation between the properties, and not on the simultaneous reality of both alternatives. The ``superposition'' of mutually exclusive alternatives is a consequence of mathematical book keeping, not of physical reality. 

To further illustrate the point, quantum ergodicity can be applied directly to obtain the ergodic probability $p(f|a)$ from the fundamental relations $p(f|m,b)$ and $p(m|a,b)$. The derivation can be given by
\begin{eqnarray}
\label{eq:cohere}
p(f|a) &=& p(f|a,b) p(a|f,b)
\nonumber \\ 
&=& \left(\sum_{m} p(f|m,b) p(m|a,b) \right)
\left(\sum_{m^\prime} p(a|m^\prime,b) p(m^\prime|f,b) \right)
\nonumber \\
&=& \sum_{m,m^\prime} (p(m|a,b)p(a|m^\prime,b))\;(p(m^\prime|f,b)p(f|m,b)).
\end{eqnarray}
The last line corresponds to the product trace of the projectors $\mid a \rangle\langle a \mid$ and $\mid f \rangle\langle f \mid$ in the Hilbert space formalism, where the products of conditional probabilities for $m$ and $m^\prime$ describe the quantum coherence between the alternative measurement results. Since self-adjoint operators can be represented as weighted sums of their projectors, it is a straightforward matter to derive the complete operator algebra of quantum mechanics from the law of quantum ergodicity, without a separate definition of state vectors. 

\section{Derivation of the Schr\"odinger equation and the physics of gauge transformations}

\label{sec:schroedinger}

The discussion above has focused on the statistical evidence obtained in measurements of quantum systems, yet quantum mechanics was originally derived from a combination of ad hoc assumptions about physical properties that were not observable with the technologies then available. In particular, the standard problem of finding the energy eigenvalues of a particle in a potential using the Schr\"odinger equation was merely motivated by the identification of transition frequencies with energy differences. The problem of electron position in the atom or the problem of its momentum seemed to be purely academic at the time. Nevertheless, this somewhat artificial problem is usually taken as the starting point of introductions to quantum mechanics. It may therefore be helpful to illustrate the relation between quantum ergodicity and the conventional formulations of quantum mechanics by applying it to this particularly familiar example. 

As the discussion in the previous section has shown, quantum states are merely a modified representation of complex conditional probabilities that describe the fundamental relations between a physical property $m$ and two other properties, $a$ and $b$. The state appears as a vector, because the law of quantum ergodicity says that the relation of $p(m|a,b)$ and $p(m|a,b^\prime)$ can be derived from the transformational distance between $b$ and $b^\prime$ encoded in the complex phases of $p(m|a,b)$ obtained for the different possible values of $m$. We can now derive the time independent Schr\'odinger equation by applying this insight about the fundamental relations between physical properties to the specific case of position $x$, energy $E$, and momentum $p$ of a single particle. 

The law of quantum ergodicity states that the relation between these three properties is given by complex conditional probabilities of the form $p(x|E,p)$. In addition, the momentum $p$ is defined so that $x$ and $p$ are canonical conjugates, which means that $x$ is the generator of a shift in $p$ and vice versa. This definition of $p$ has two important consequences. Firstly, it means that the ergodic probabilities $p(x|p)$ are constant, since a completely random shift in momentum means that every final momentum $p$ has the same probability. Secondly, it is possible to identify the translational distance $d_{(p|x)}$ between $p$ and $E$ along $x$ with a momentum difference given by the gradient of the phase of $p(x|E,p)$ in $x$. 

To find the correct quantum mechanical expression for the relation between position, energy and momentum, we need to modify the classical function $E(x,p)$ so that the difference between the momentum $p$ in $p(x|E,p)$ and the classical momentum obtained for $x$ and $E$ in the classical relation approximately corresponds to the transformational distance of quantum ergodicity. This can be achieved by ``correcting'' the momentum $p$ in the quantum ergodic relation $p(x|E,p)$ by a derivative in $x$ that extracts the phase gradient at $x$ from $p(x|E,p)$ and thus provides a mathematical definition of the translational distance along $x$,
\begin{equation}
\label{eq:pdistance}
d_{(p|x)} = -i \hbar \frac{\partial}{\partial x}.
\end{equation}
With this ``correction'' of the momentum $p$ in the relation $p(x|E,p)$, we can translate the classical relation into its proper quantum form, following a procedure that is indeed reminiscent of the axiomatic replacement of momentum with an operator in the traditional approach. The difference is that this replacement is now motivated by a more general law that governs all deterministic relations between physical properties. 

In the non-relativistic case, the complex conditional probabilities $p(x|E,p)$ that define the correct quantum mechanical relation between position, energy and momentum can be derived from the quantum ergodic form of the Schr\"odinger equation,
\begin{equation}
\label{eq:schroedinger}
\frac{\left(-i \hbar \frac{\partial}{\partial x} + p \right)^2}{2 m} p(x|E,p) + V(x) p(x|E,p) = E p(x|E,p).
\end{equation}
For a reference momentum of $p=0$, this is the standard form of the time independent Schr\"odinger equation, where the complex conditional probability is related to the wavefunction by the normalization factor given in Eq.(\ref{eq:amp}) above. However, the quantum ergodic relations are more complete than the wavefunction, because they replace the seemingly arbitrary phases of $\psi(x)$ with a well defined relation between $(x,E)$ and the reference momentum $p$. This means that quantum ergodicity provides a proper explanation of gauge transformations: a difference in gauge simply corresponds to a different choice of reference $p$. Specifically, the reference should be defined in terms of physical properties. In the conventional case, momentum is proportional to velocity and $p=0$ means that the particle is at rest in this frame of reference. 

The appearance of $p$ in Eq.(\ref{eq:schroedinger}) corresponds to the most simple gauge transformation, where the reference state $p$ represents a constant velocity different from zero. In the presence of three dimensional gauge fields, it may not be possible to find a state of constant velocity $v$ that is also a canonical conjugate to position, because the gauge field introduces ergodic probabilities of the form $p(v_x,v_y)$ that dynamically relate the components of velocity to each other. To obtain an unbiased ergodic probability for this case, the reference $p=0$ must be defined as a specific combination of three dimensional positions and velocities, such that the $p$-dependent term in Eq.(\ref{eq:schroedinger}) is replaced by an appropriate spatial dependence of the vector potential. 

In the present field-free case, it is also possible to apply gauge transformations, either by shifting the reference velocity, or by applying the general transformation given in Eq.(\ref{eq:transform}). These gauge transformations illustrate the fact that $\psi(x)$ is actually a differently normalized form of the complex conditional probability $p(x|E,p)$, where the phases are determined by the reference $p$. The axiomatic definition of $p$ using the operator of transformational distance of Eq.(\ref{eq:pdistance}), which is typically used in conventional quantum mechanics, is not sufficient to properly identify the physical meaning of $p$, since these can only be known if the actual observable properties associated with $p=0$ are defined as well. 

Quantum ergodicity shows that the momentum reference must be included to describe the complete physics of the state, since the phases of $\psi(x)$ are really determined with respect to a reference $p$ in the corresponding complex conditional probability $p(x|E,p)$. The general form of gauge transformations between different conjugate references is given by Eq.(\ref{eq:transform}), where it is shown that a transformation of the reference $b$ along $m$ corresponds to a phase change in the complex conditional probabilities. For a shift in reference momentum from $p^\prime$ to $p$, 
\begin{equation}
\label{eq:pshift}
p(x|E,p) = \frac{p(x|E,p^\prime) \exp(\frac{i}{\hbar}(p^\prime - p) x)}{\int p(x^\prime|E,p^\prime) \exp(\frac{i}{\hbar}(p^\prime - p) x^\prime) d x^\prime}.
\end{equation}
This relation means that we can derive the complex conditional probabilities $p(x|E,p)$ for all $p$ from only a single reference $p^\prime$. We can then derive the transformation between the position representation $p(x|E,p)$ for a single reference momentum $p$ and the momentum representation $p(p|E,x)$ for a single reference position $x$ using the relation
\begin{equation}
\label{eq:conjugate}
p(x|E,p) p(p|E,x) = \frac{1}{2 \pi \hbar}
\end{equation}
which is the law of quantum ergodicity for the special case of canonical conjugation, where $p(x|p)=1/(2 \pi \hbar)$. By combining Eq.(\ref{eq:pshift}) with Eq.(\ref{eq:conjugate}), we obtain the relation between the representations as
\begin{equation}
\label{eq:FT}
p(p|E,x) = \frac{\int p(x^\prime|E,p^\prime) \exp(\frac{i}{\hbar}(p^\prime - p) x^\prime) d x^\prime}{2 \pi \hbar p(x|E,p^\prime) \exp(\frac{i}{\hbar}(p^\prime - p) x)}.
\end{equation}
For references of $p^\prime=0$ and $x=0$, this corresponds to the Fourier transform relation between the wavefunction in the position representation and the wavefunction in the momentum representation. Note however that the normalization of the complex probabilities requires an additional factor proportional to the conditional probability at the reference point. 

In principle, the analysis above can be extended to cover the whole range of problems covered in conventional quantum mechanics, including quantized fields and relativistic particles. In fact, it is not even necessary or desirable to focus on Hamiltonian formulations of physics. The law of quantum ergodicity can be applied directly to any deterministic relation between the physical properties of a system, e.g. to Newtonian or relativistic laws of motion. It is therefore a much more flexible ``law of quantization'' than any of the previously known procedures. This might be a crucial advantage in situations where Hamiltonian or Lagrangian approaches are difficult to apply, e.g. in quantum gravity. It may therefore be worthwhile to reflect a bit more on the differences between the original formulation of quantum mechanics and the fundamental physics described by the law of quantum ergodicity.

\section{Criticism of established concepts}

In the light of the present results, it seems that the concepts of ``operators'' and ``states'' introduced in the original formulation of quantum mechanics are completely dispensable and may actually have distorted our view of quantum physics. Since this is a rather disturbing thought, it may be necessary to address it directly. Historically, the notion of states emerged from Bohr's model of the atom, where it was simply postulated that the experimentally inaccessible situation inside the atom could be summarized in this manner. The only connection to the actual physics was provided by the well-defined energy, and this was later developed into the notion of ``eigenstates'', where one physical property is known with precision, while the others appear to be random. In the Hilbert space formalism, this notion is used to separate the description of physical properties from the description of ``states'' by introducing the concept of ``operators''. The operator algebra can express all relations between physical properties, but the experimental evidence can only be explained in terms of the statistics of a specific state. The operator algebra of Hilbert space thus suggests an odd kind of dualism between universal laws of physics and the individual measurement results obtained under well-defined circumstances. 

Quantum ergodicity resolves this problem by unifying ``states'' and ``operators'' in terms of universal relations between physical properties. Note that these universal relations contain no randomness. Instead, they replace the laws of physics previously given in terms of functions directly relating the values of observables to each other. For example, the classical limit of the Schr\"odinger equation is simply given by the Hamiltonian relating energy to position and momentum,
\begin{equation}
\label{eq:classic}
E = H(x,p).
\end{equation}
In the classical limit, the approximate relation between energy, position and momentum would be given by
\begin{equation}
\label{eq:cprox}
p(E|x,p) \approx \delta(E-H(x,p)).
\end{equation}
However, the correct expression needs to obey the law of quantum ergodicity and is therefore given by a complex conditional probability that satisfies the relation
\begin{equation}
\label{eq:Hergodic}
p(E|x,p) p(x|E,p) = p(x|E).
\end{equation}
Therefore, the actual conditional probabilities $p(E|x,p)$ are complex, where the gradient of the complex phase represents the transformation distance between the properties $E$, $x$, and $p$. The classical approximation given by Eq.(\ref{eq:cprox}) only applies when the probabilies are coarse grained, so that rapidly oscillating phases result in probabilies of zero, leaving only a probability of one around the classical result given by $H(x,p)$. 

Significantly, these results mean that all classical relations of the form $H(x,p)$ are approximations that should be replaced by the more fundamental relations given by $p(E|x,p)$ and $p(x|E,p)$. As shown in section \ref{sec:schroedinger}, the wavefunction of an energy eigenstate actually represents these fundamental relations, irrespective of the state of a system. The historic misunderstanding that the wavefunction should be identified with the ``state'' of a particle arises from the fact that the fundamental relations between $E$, $x$, and $p$ also determine the ergodic probabilities $p(x|E)$. In fact, the correct explanation of the physics is that the wavefunction of an eigenstate of $E$ is merely a renormalized expression of the relation between energy, position and momentum for the reference momentum $p=0$,
\begin{equation}
\label{eq:wave}
\psi_E(x)=\sqrt{\frac{p(p=0|E)}{p(x|p=0)}}p(x|E,p=0).
\end{equation}
The reason why this fundamental relation between energy, position and momentum can be used to predict all ergodic probabilities for a system prepared in a state with well defined energy is that the transformation 
laws of Bayesian probabilities given by Eq.(\ref{eq:chain}) define an inner product that can be combined with the law of quantum ergodicity to obtain the conventional formula for quantum probabilities also known as Born's rule. 

The confusion about the meaning of the wavefunction originates from the mistaken assumption that it describes the statistics of a specific situation rather than a fundamental relation between physical properties. In this paper, I have shown that this is not correct. When properly indentified in terms of empirical concepts and procedures, the algebra of quantum mechanics originates from complex conditional probability that describe the correct quantum limit of the deterministic relations between physical properties. As shown above, the classical relations are merely an approximation of these complex conditional probabilities, where a probablity of one is assigned to transformational distances of zero while the probabilities of all other values are neglected. 

\section{The relation between universal laws and statistical evidence}

Much confusion originates from the problem that the experimental evidence obtained from quantum systems is necessarily statistical. It is therefore important to understand how the familiar statistical patterns observed in specific quantum measurements emerge from the fundamental relations of quantum ergodicity. For this purpose, the actual quantum state describing the statistics of a specific situation should be expressed in terms of a joint probability $\rho(a,b)$ referring to a complementary pair of  observable properties, $a$ and $b$. As the recent experimental evidence shows, this complex joint probability is the one directly obtained from  weak measurements of $a$ followed by a precise measurement of $b$ \cite{Lun12}. It is then possible to determine the probability of any measurement result $m$ by applying the conventional Bayesian rules to the joint probability $\rho(a,b)$ and the universal relation between $a$, $b$ and $m$ given by the conditional probability $p(m|a,b)$ \cite{Hof12},
\begin{equation}
p(m) = \sum_{a,b} p(m|a,b) \rho(a,b).
\end{equation}
Here, the relation between previous information and future prediction describes the fundamental physics. In the original formulation of quantum mechanics, a serious misunderstanding arose because the discussion focused only on predictions from ``pure'' states, resulting in the mistaken conclusion that such states should be fundamental elements of reality. However, ``pure'' states simply represent situations were one property is known with precision, while all others are randomly distributed according to quantum ergodicity. For an initial state with known $m$, the joint probability is then given by
\begin{eqnarray}
\rho(a,b|m) &=& p(a|m,b) p(b|m)
\nonumber \\ &=& p(b|a,m) p(a|m)
\nonumber \\ &=& p^*(m|a,b) p(b|a).
\end{eqnarray}
Thus, quantum ergodicity does result in a fundamental connection between the universal laws of physics expressed by complex conditional probabilities and the observable statistics of pure states. However, this connection has been misinterpreted due to the use of a vocabulary borrowed from classical wave theory, where the additions of complex probabilities are misinterpreted as ``interferences'' and the essential role of the third property is overlooked. 


The law of quantum ergodicity provides a consistent explanation of quantum mechanics based on universal laws of physics that do not depend on the specific situation. With this new foundation, it is possible to revisit all of the scenarios described by conventional quantum mechanics. The most significant change is that ``superposition'' now expresses the relation between different possible realities that can never occur jointly. For instance, the double slit problem is now described as a relation between which-path measurements $x$ and measurements of momentum $p$ for a well-defined double slit property $\psi$ that relates the two to each other according to the complex conditional probability $p(\psi|x,p)$. The interference pattern is merely the ergodic distribution $p(p|\psi)$ of momentum for that double slit property, and its measurement limits the effects of the particle to a reality defined by the set of properties $(\psi,p)$, a reality that is physically distinct from the which-path reality of $(\psi,x)$. Importantly, physical reality requires interaction, and the interaction associated with a which-path measurement is incompatible with the alternative measurement of momentum.  

Interestingly, this line of argument has been used from the beginning of quantum mechanics. However, it has not been properly connected to the mathematical formulation. Nothing much was done to provide useful alternatives to the misconception that ``superpositions'' somehow describe simultaneous realities. The idea that one should simply avoid any reference to unobserved properties has opened the doors to wild speculations about ``realities'' beyond all experimental observations. However, the mathematical structure of quantum mechanics does permit much clearer statements about the physics. In the end, the only consistent interpretation of the observable results is that reality only emerges in interactions, and that there is no static reality in the microscopic limit, where the effects of the necessary interactions cannot be neglected anymore. The level of interaction where the separation of dynamics and reality is valid finds its quantitative expression in the action-phase ratio $\hbar$, which explains why the notion of a measurement independent reality is a good approximation at the macroscopic level. The law of quantum ergodicity thus provides a clear quantitative description of the inseparable relation between dynamics and reality that is at the heart of quantum mechanics, and finally achieves a reconciliation of the fundamental formulation of physics with its classical limit. 

\section{Conclusions}

The present paper is the starting point for an extensive revision of quantum mechanics. The discussion above shows that quantum mechanics can be explained completely without any mathematical assumptions such as state vectors or operators. Instead, the law of quantum ergodicity is a well defined modification of the relation between experimentally observable properties. This relation is itself is based on the experimental evidence obtained in weak measurements. It is therefore not obtained from mathematical speculations or invented theories, but a necessary consequence of experimental observations. In the future, introductions to quantum mechanics could therefore be based on directly observed phenomena, proceeding from physical evidence to mathematical descriptions without the need to ``shut up and calculate''.

A significant consequence of the law of quantum ergodicity is that it provides the proper expression for fundamental laws of physics. In the original formulation of quantum mechanics, laws of motion were replaced by operator equations, leaving the relation to individual systems unclear. Likewise, the evolution of the state vector merely described the time dependence of averages, not the dynamics of individual systems. The law of quantum ergodicity shows that the intrinsic time evolution of a system has no physical reality, because properties observed at different times are related by complex conditional probabilities that express the dynamics of the system in terms of action phases. This means that the laws of motion are really given by complex conditional probabilities, while the idea of time dependence as a continuous trajectory is merely an approximation. It is in fact wrong to think of physical objects as geometric shapes in space and time. Instead, we need to realize that the experimental evidence of reality is given by the gradual emergence of interaction effects represented by quantum ergodic probabilities. 

Ultimately, the identification of universal laws of causality using quantum ergodicity will have far reaching consequences, since it redefines the relation of quantum mechanics with all other branches of physics and places the results of quantum physics into a much larger context. I realize that the revision of quantum mechanics required by this insight is quite a challenge, and it might be tempting to hold on to the familiar form we all learned from our textbooks. However, we should not forget the confusion that the original formulation of quantum theory has caused in our understanding of physics and of the world around us. Many of the recent results in quantum optics and quantum information appear to be paradoxical and counter intuitive, and there are bitter disagreements regarding the interpretation of the present formalism. In the light of the present results, it seems that this confusion is the consequence of a historic misunderstanding created by the unfortunate choice of problems, which were not dominated by measurement, but by speculations about static realities inside atoms that were completely inaccessible to experiment. It may well be that all of the interpretational problems of quantum mechanics merely arose because of this historic limitation to the wrong set of problems. 

The discovery of quantum ergodicity is a natural consequence of the great advances in experimental methods that have enabled us to finally control individual quantum systems with optimal precision. It is firmly based on the new experimental evidence that has become available as a result of the admirable efforts of researchers exploring phenomena at the very edge of our understanding. In the tradition of science, we should therefore be ready to leave preconceived notions behind and follow the evidence wherever it may lead us. 

\section*{Acknowledgment}
I would like to thank Phillipe Grangier, Mike G. Raymer, Noboyuki Imoto, Lorenzo Maccone and Masataka Iinuma for interesting and motivating comments. This work was supported by JSPS KAKENHI Grant Number 24540427.

\end{document}